\documentclass[aps,floatfix,11pt]{revtex4}
\usepackage{graphicx}
\begin{document}
\title{Cosmic Rays at the Knee}
\author{Thomas K. Gaisser}
\affiliation{Bartol Research Institute and Department of Physics and Astronomy,\\
University of Delaware,
Newark, DE 19716}

\begin{abstract}
Several kinds of measurements are
combined in an attempt to obtain a consistent estimate of the spectrum
and composition of the primary cosmic radiation through the knee region.
Assuming that the knee is a signal of the high-energy end of a
galactic cosmic-ray population, I discuss possible signatures of a
transition to an extra-galactic population and how they might be detected.  
\end{abstract}

\maketitle

\section{Introduction}\label{sec1}

\begin{figure}
\begin{center}
\includegraphics[width=13cm]{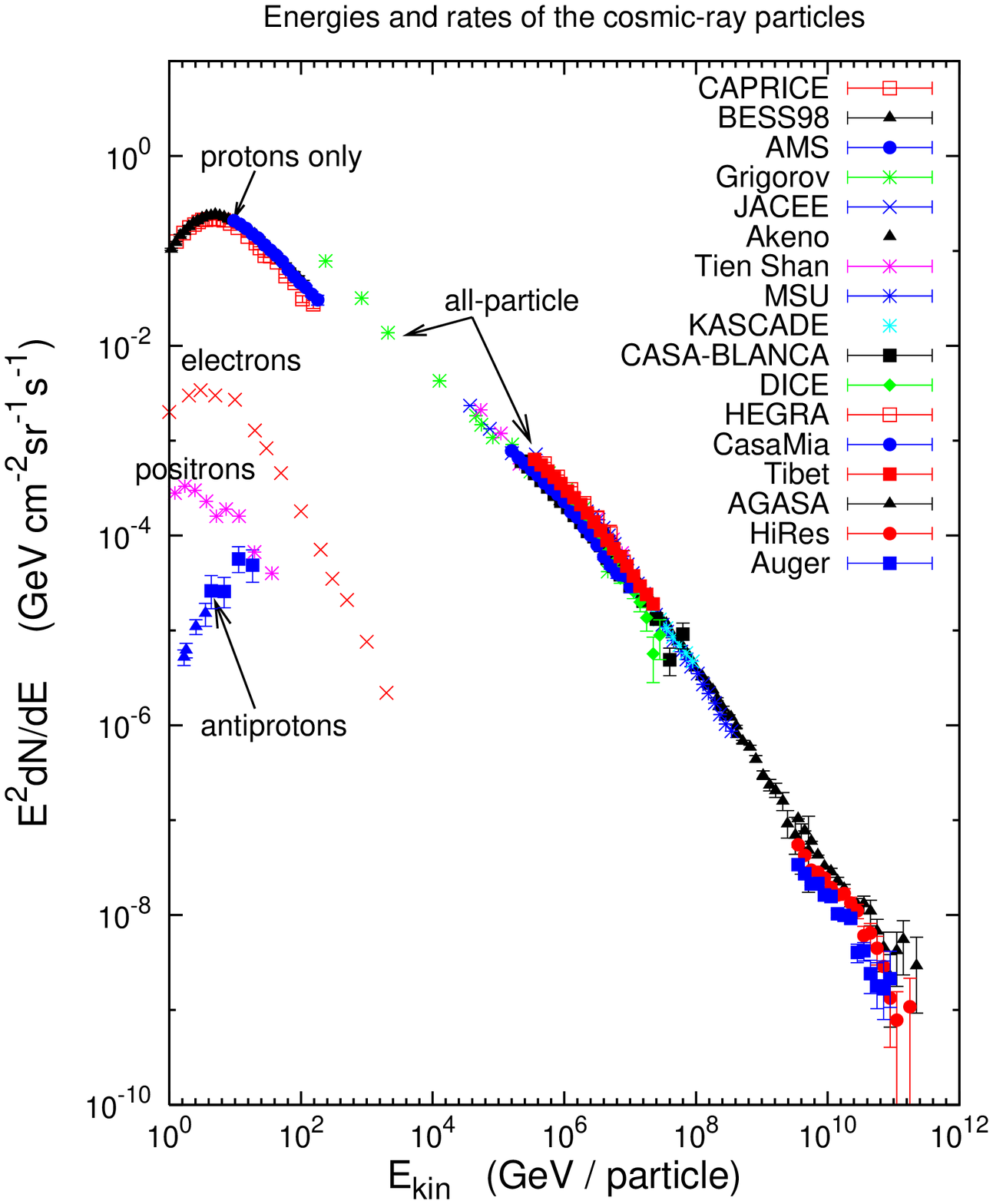}

\end{center}
\caption{Spectrum of cosmic rays}
\label{fig1}
\end{figure}

The cosmic-ray spectrum extends from the sub-GeV region to at least
$10^{11}$~GeV (Fig.~\ref{fig1}.)  Up to $100$~GeV and somewhat higher,
measurements with magnetic spectrometers flown above most of
the atmosphere provide good momentum resolution along with identification of 
the charge and mass of individual primaries.   Measurements with
calorimeters continue to identify individual primaries to beyond $100$~TeV,
but with larger systematic uncertainties in the energy assignment.
In the PeV region and beyond, the cosmic-ray intensity is too low
for direct measurements; only indirect measurements of air showers
from the ground are possible.  Since the particles are not identified
on an event-by-event basis, the energy spectrum derived from
measurements of air showers is given as an
``all-particle" spectrum, in terms of energy per particle rather than
energy per nucleon.  In the air-shower regime, identification of the
primary mass is made in one of several indirect ways on a statistical basis,
complicating the search for features in spectra of individual elements.

The gyroradius of a proton in a typical galactic magnetic field of
3~$\mu$Gauss is about half a parsec at one PeV.   The parsec scale is also typical
of the size of structures in the interstellar medium driven by supernova
explosions.  For particles with higher energy and
larger gyroradii, diffusion in the interstellar medium may become
less efficient.  Estimates of the maximum energy of particles accelerated
at supernova shocks are around the same energy.  
It is therefore natural to guess that the knee of the cosmic-ray
spectrum around 3 PeV has something to do with the end (or at least
the beginning of the end) of the galactic cosmic-ray spectrum.

Peters~\cite{Peters} described the consequences for energy dependence
of the primary composition if the spectrum is characterized by
a maximum rigidity, $R_c$, which could be associated either with propagation
or with acceleration (or both).  The relation between rigidity
and total energy is
\begin{equation}
R\;=\;{P\,c\over Z\,e},
\end{equation}
where $P\,c\approx E_{\rm tot}$ is the total energy of a nucleus of
charge $Z$ and mass $A$.  If cosmic rays are classified by energy per particle,
as is the case for air shower measurements, then the spectrum should steepen
first for protons, then for helium, then for the the CNO group etc.
Elsewhere~\cite{erix} I have called this sequence the "Peters cycle".  Several 
air shower measurements show some evidence that the spectrum becomes
progressively enriched in heavy nuclei through the knee region.  The
clearest evidence for the Peters sequence for several groups of nuclei
comes from analysis of the KASCADE experiment~\cite{KASCADE}.

\section{Comparison of direct and indirect measurements}\label{sec2}
 Data from direct measurements above 100 TeV are sparse,
leaving a gap that is bridged by emulsion chamber data
with low statistics below a PeV and by the threshold region of
air shower experiments above 100 TeV.  Apart from the technical
problems of low statistics and systematic threshold effects,
there is also the problem that air shower experiments do not 
identify individual primary nuclei.  In addition, the efficiency
of ground arrays depends strongly on primary mass in the threshold region.
Nevertheless, it is possible to form a fairly consistent picture
of the cosmic-ray spectrum up to the knee.

\begin{figure}
\begin{center}
\includegraphics[width=7cm]{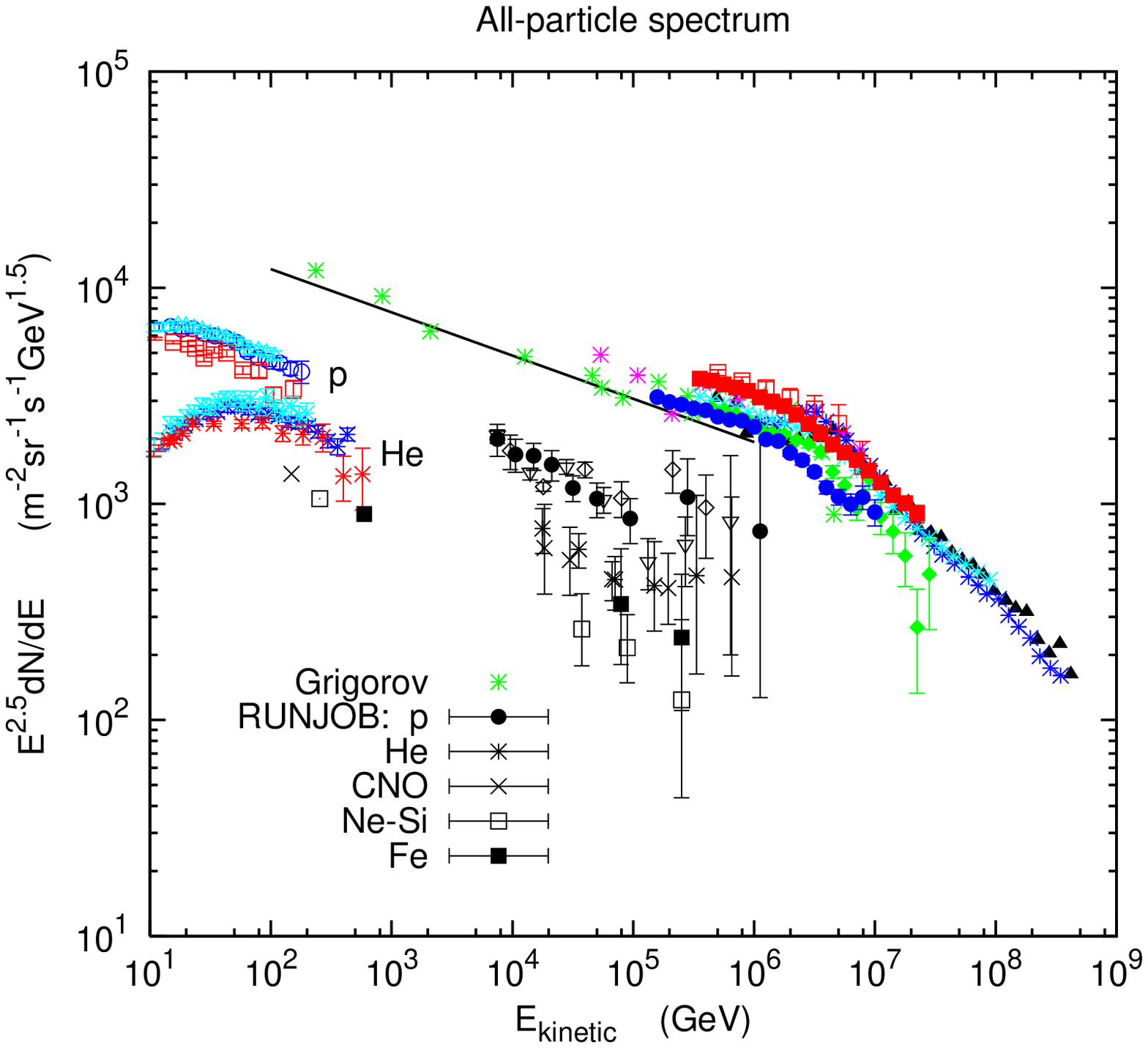}
\includegraphics[width=7cm]{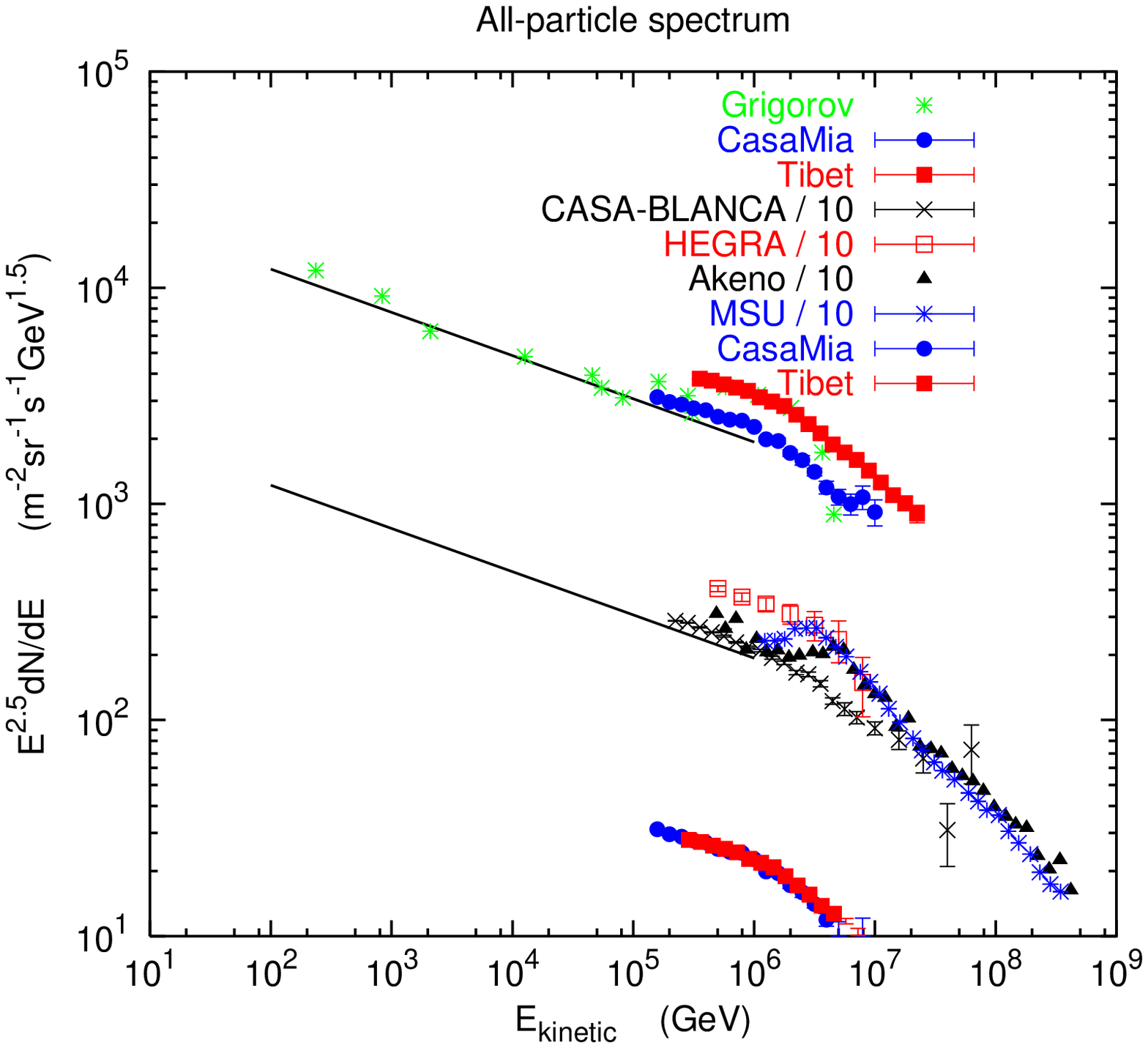}
\end{center}
\caption{All-particle spectrum.  Upper panel illustrates an extrapolation
from direct measurements at low energy to 1 PeV.  Lower panel shows detail
of air shower measurements in the knee region.  The two lower sets of points
are offset respectively by $1/10$ and by $1/100$ for clarity (see text).
}
\label{fig2}
\end{figure}

Figure~\ref{fig2} shows several measurements of the primary spectrum
through the knee region.
Up to 100 GeV there are good measurements of the spectra of protons and
helium with magnetic spectrometers flown in spacecraft~\cite{AMS}
and high-altitude balloons~\cite{BESS,CAPRICE}.  
Using the AMS~\cite{AMS} and BESS~\cite{BESS}
measurements, together with measurements of heavier nuclei 
at $10$~GeV/nucleon~\cite{HEAO3}, I have converted the spectra of protons, helium,
CNO, Ne-Si and Fe from energy per nucleon to total energy per particle and
combined them to give the all particle spectrum at low energy (shown as the solid line
in Fig.~\ref{fig2}).

The highest energy direct measurements in which individual nuclei are 
identified directly are from emulsion chamber experiments.  Measurements
of protons and helium from RUNJOB~\cite{RUNJOB} and JACEE~\cite{JACEE}
are shown in Fig.~\ref{fig2}a.  Preliminary data
on protons and helium from the ATIC thin ionization
chamber~\cite{ATIC} (not shown here) are consistent with a smooth power-law extrapolaton
between the BESS and AMS spectromter data at low energy and the RUNJOB data
above $10$~TeV.  (The JACEE helium data are higher.)  The solid line normalized
to the all-particle spectrum at 100 GeV and extrapolated to 1 PeV
with an $E^{-2.7}$ power law is consistent with the all-particle measurements
of Grigorov~\cite{Grigorov} but about a factor 1.5 above the sum of the RUNJOB
spectra at 100 TeV.  (Using the higher JACEE measurement of helium would bring the
emulsion chamber measurements into agreement with the $E^{-2.7}$ extrapolation
at 100 TeV.)  An external motivation for using the hard extrapolation
from the spectrometer measurements comes from the Super-K measurement of the
flux of atmospheric neutrinos, where the best fit to the data requires 
an even harder extrapolation of the primary spectrum above 100 GeV~\cite{Super-K}.

Even with the hard $E^{-2.7}$ spectrum, the extrapolation of the direct
measurements comes in somewhat below the air shower measurements. 
Moreover, the all-particle spectrum of RUNJOB~\cite{RUNJOB2} is 
somewhat below the lowest-energy air shower data, so there
may be some systematic offset between the direct measurements and the
measurements of air showers.  There is also a slight correlation between
hardness of the fitted spectrum and primary mass in this energy range~\cite{RUNJOB3}.
There is, however, no sign \cite{RUNJOB4} of the sharp steepening of the proton spectrum,
below $100$~TeV,
which would be the case if the maximum rigidity
accessible in most galactic cosmic-ray accelerators were around $100$~TV
or below as originally estimated for diffusive shock
acceleration by expanding supernova remnants~\cite{LagageCesarsky}.

\section{Modelling the knee}\label{sec3}

Although there are systematic differences among the various air shower 
measurements in the knee region, all show that the spectrum steepens
from $-2.7$ or slightly harder below $1$~PeV to $-3$ above $10$~PeV.
Figure~\ref{fig2}b shows several results. Two of the 
measurements~\cite{Akeno,MSU} show evidence of structure in the
knee region, while the others~\cite{CASA-M,Tibet,HEGRA,CASA-B} do not.
The figure also shows the spectra of two of the measurements (CASA-MIA and Tibet)
with the energy assignments of the latter shifted down by a factor
of $0.8$, which may be taken as an indication of systematic uncertainties
in energy assignment.  The shapes of these two spectra agree well with each
other.

Some authors~\cite{EW} have suggested that a single source contributes
significantly to the flux of cosmic-rays in the knee region.  The
possible offset between the air-shower measurements and the extrapolation
of the direct measurements leaves room for such a possibility.  
Others~\cite{Axford,Volk}
argue, however, that the overall smoothness of the spectrum indicates
a single population, with a secondary acceleration mechanism boosting
some of the cosmic rays accelerated by galactic supernova explosions to 
higher energy.  Several authors~\cite{Bergman,Berezinsky,Hillas} approach 
the problem from the high-energy end by
modelling the highest energy spectrum as arising from a cosmological distribution
of sources and subtracting the extra-galactic population from a model of
the supernova-accelerated population to see if an extra, high-energy galactic
source is required to fill in the energy region above the knee (``population B"~\cite{Hillas}).
This is an open question at present.  
(For a recent review and further references see ~\cite{Hillas-b}.)

In any case it is interesting to estimate the power required to produce the
high-energy end of the galactic cosmic-ray spectrum.  To do so, I assume
a model of galactic propagation with diffusion characterized by
an equivalent leaky box model with a escape time given by
\begin{equation}
\tau_{\rm esc}\;=\;2\times 10^7\,{\rm yrs}\times E^{-0.33}
\end{equation}
which applies for all energies.  With this propagation model (motivated
by lack of anisotropy at high enery), the source spectrum at low energy
must be $E^{-2.37}$ to give the observed $E^{-2.7}$ spectrum inside the galaxy.
Assuming a maximum rigidity of $1$~PV for this low-energy component, with
a composition as measured by an emulsion chamber experiment~\cite{RUNJOB},
the assumed low-energy component can be subtracted from the observed spectrum.
The total power required to account for the observed spectrum up to $1$~EeV
is then $\sim 2\times 10^{39}$ erg/sec.  Such an estimate is obviousy very
model dependent, but not unreasonably large compared to what might be
available in individual galactic sources.

\section{Primary composition from air shower measurements}\label{sec4}

What is needed to make progress is a precise knowledge of the energy-dependence
of the major groups of nuclei (p, He, CNO, heavy) through the knee region.
If there are several important groups of sources with different maximum
rigidities then there should be a corresponding sequence of Peters cycles,
perhaps characterized by different compositions.  The transition to
extra-galactic cosmic-rays would be characterized by a transition from
heavy nuclei (from the highest galactic source) to the light component of
a cosmological distribution of sources.  A recent summary of composition measurements
with air showers in the knee region is given in the rapporteur paper of
Matthews at the last ICRC~\cite{Matthews} (See also Swordy {\it et al.}~\cite{Swordy}.)  
Overall, the evidence
suggests a change in composition toward hearvier primares at higher energy
as expected.  However, results of different measurements
disagree in detail and the picture is not very clear.  As noted in the
introduction, the best evidence for the sequence of increasingly heavier
groups of nuclei comes from the KASCADE experiment~\cite{KASCADE}, which uses the
ratio of muons to electrons as a probe of primary composition.

Ratio of muons to electrons in the shower front is a classic probe of primary
composition.  At each interaction of the nucleons in an air shower,
roughly 1/3 of the energy not retained by the projectile nucleon
is transferred to the electromagnetic component of the cascade
via $\pi^0\rightarrow \gamma\,\gamma$ and 2/3 to
charged pions.  The charged pions either reinteract or decay depending
on their energy and the depth (and hence the density) in the atmosphere where they 
are produced.  Charged pions that interact
contribute further to the electromagnetic component while those that decay
feed the muon component of the air shower.  Comparing nuclei of the
same total energy, charged pions reach the energy
at which they can decay earlier in the cascade for heavy primaries 
than for protons because the initial energy per nucleon is lower by 
$E_{\rm total}/A$.  As a consequence, the ratio of the muonic to the
electromagnetic component of an air shower is larger for heavy primaries.

To measure this ratio with greatest sensitivity requires an observation near
shower maximum so that the size of the shower is well correlated with the
primary energy.   In general, showers in the knee region and somewhat above
are observed after shower maximum so that fluctuations in the relation between
observed shower size and primary energy are important.
KASCADE is a surface array near sea level so that these fluctuations are large.

\begin{figure}[thb]
\begin{center}
\includegraphics[width=7cm]{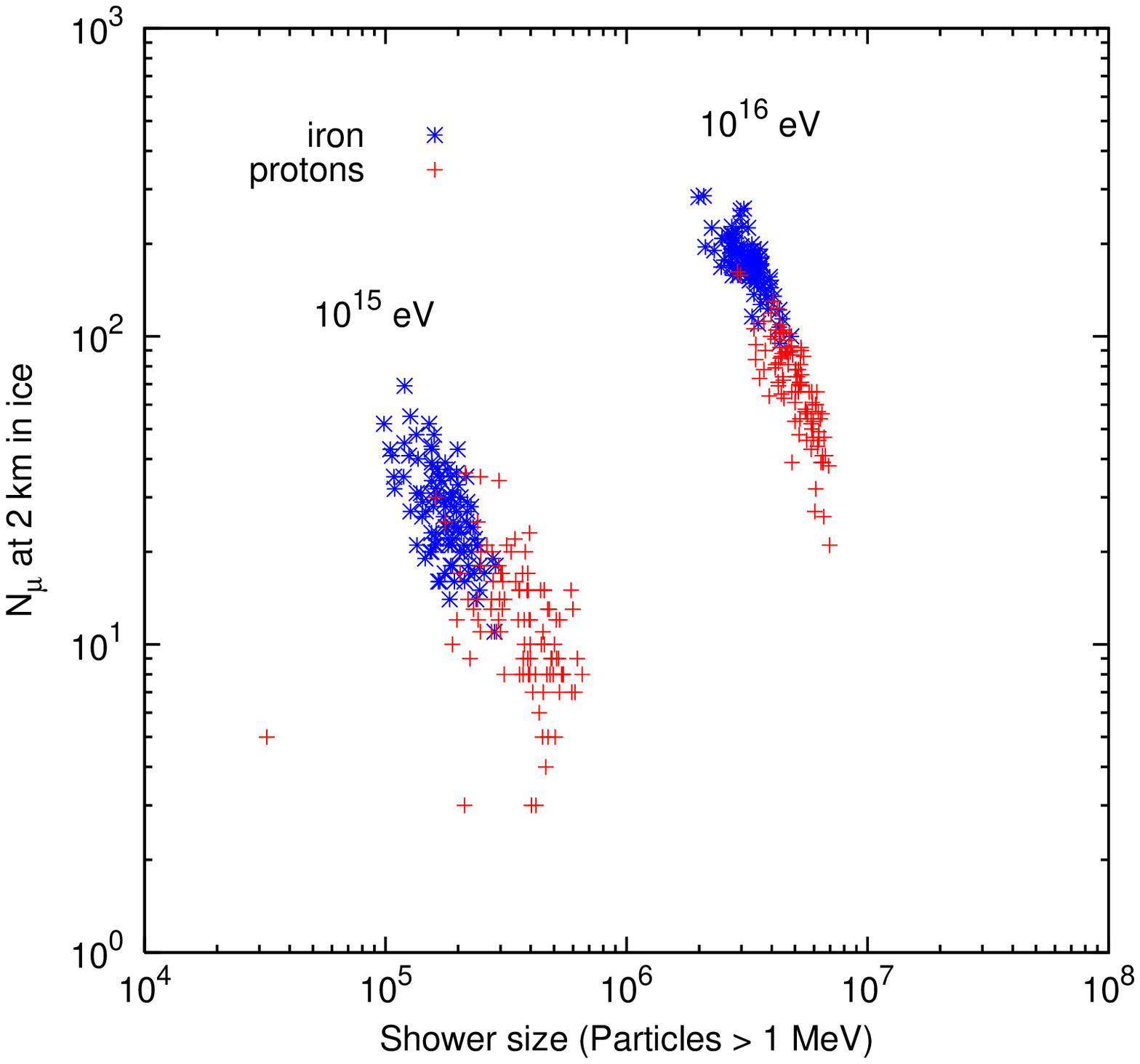}
\includegraphics[width=7cm]{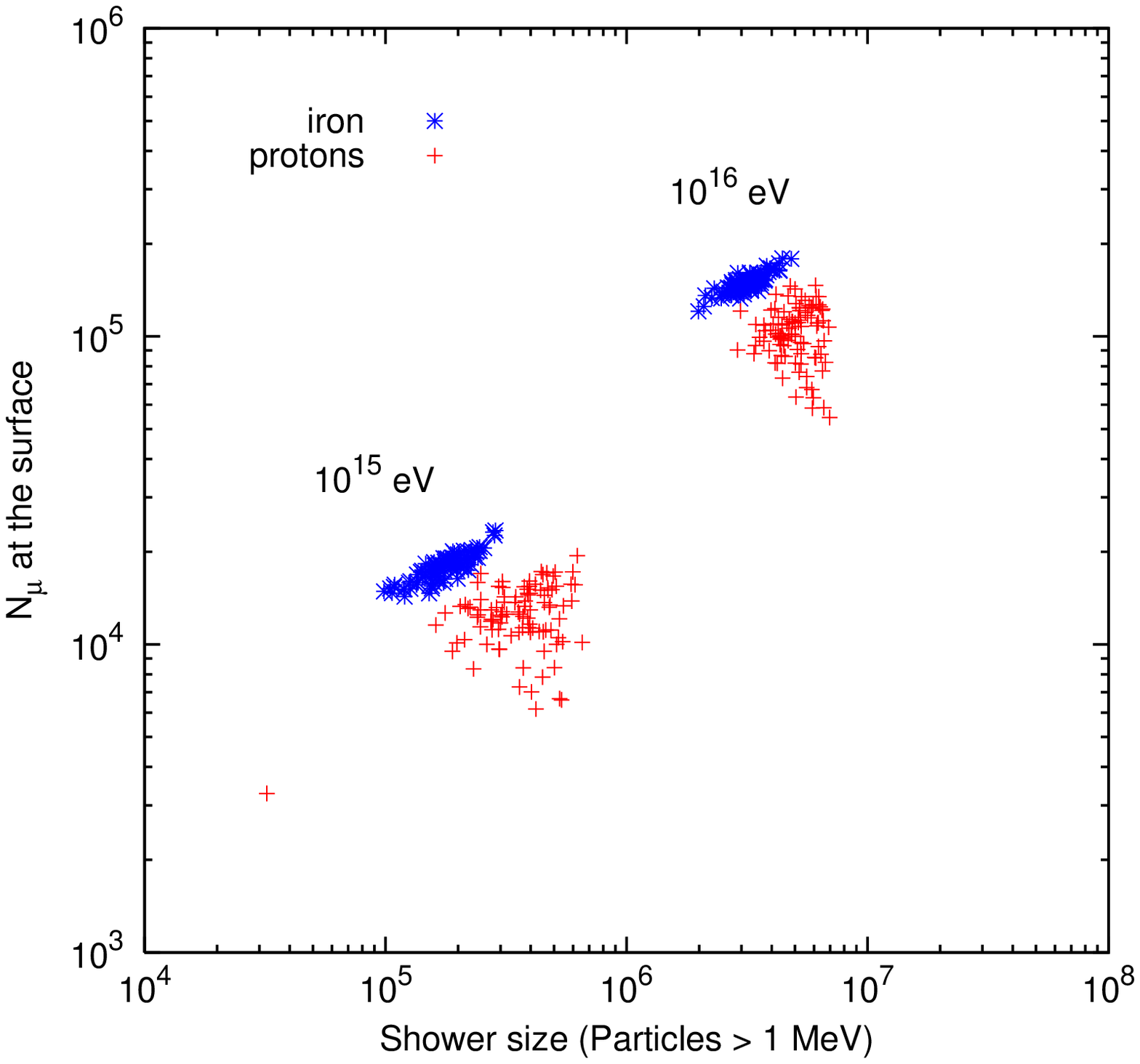}
\end{center}
\caption{Correlation between muons and shower size at the surface.  Left panel,
muons at 2 km in the ice; right panel, muons at the surface.
}
\label{fig3}
\end{figure}

Muons detected in KASCADE typically have energies of a few GeV.
An alternative, realized with EASTOP-MACRO~\cite{E-M} and with SPASE-AMANDA~\cite{SPAM},
is to sample the muon component with a deep underground detector and the
electromagnetic component with an array on the surface above.  Such a setup
selects muons with sufficiently high energy at production to penetrate to the 
deep detector.  The high-energy
muons generally come more from the fragmentation region of phase space of the 
hadronic interactions in the shower, whereas than the low-energy muons reflect
more the less well-understood central region.  On the other hand,
the multiplicity of high-energy muons is small so that fluctuations
in muon number are more important.  Some of these points are illustrated
in Fig.~\ref{fig3} which compares $N_\mu$ vs. $N_e$ for high-energy ($>0.5$~TeV)
muons and for low-energy muons in $10^{15}$ and
$10^{16}$ eV showers simulated with CORSIKA.  In general, the mass resolution
is somewhat better for the low energy muons provided that the sampling is
good enough to get a good measure of the muon number in individual showers,
while the energy resolution is better for coincident events with high energy
muons, particularly if the surface array is at high altitude.  This
complementarity was pointed out by Ralph Engel~\cite{Engel}.

It is interesting that analyses of both EASTOP-MACRO~\cite{E-M} and 
SPASE-AMANDA~\cite{SPAM} suggest an increase in the mean primary mass
in the decade of energy above a PeV, that is, through the knee region.
As a muon detector, MACRO has the advantage of resolving muon tracks, whereas
AMANDA reconstructs the light pool generated by the bundle, which
gives a measure of the energy deposition of the muons.
A limitation of both experiments is the relatively small sampling area of
the deep muon detectors, which has to be accounted for in the
comparison between data and simulations.  

IceCube, with its surface component
IceTop, now under construction at the South Pole~\cite{IceCube} 
will have a much larger acceptance.  Engineering data from the
first year of operation of IceCube with four IceTop stations and
one string demonstrate the ability of IceCube to reconstruct
events with few nanosecond accuracy over distances of more than
two kilometers~\cite{performance}.  
Since January 2006 IceCube has been operating with 16 surface
stations and 9 strings of detectors in the ice.  Each string contains
60 digital optical modules (DOMs) evenly spaced 
in the clear ice between 1450 and 2450 meters below the surface.
Each IceTop station consists of two ice Cherenkov tanks
separate from each other by 10 meters and located 25 m from the top
of the corresponding IceCube string.  Each tank is instrumented
with two of the same DOMs used in the ice.  With a nominal grid spacing
of $125$~m, the acceptance of the current partial array is
$\sim\,1500$~m$^2$sr, which allows detection of coincident events
approaching $10^{17}$~eV.  The acceptance of the full IceCube as
a three-dimensional air shower array will be approximately 
$0.3$~km$^2$sr, allowing measurement of coincident events up to
an EeV.  It is scheduled for completion in 2011 and will be operating
in the meantime as new detectors are added.

\section{Transition to extra-galactic cosmic rays}\label{sec5}

Analysis of measurements of energy-dependence of the
depth of shower maximum with HiRes~\cite{HiRes} suggest a
change in composition from heavy to light as energy increases
from $10^{17}$ to $10^{18}$~eV.  
Shower maximum is deeper in the atmosphere for protons than for
heavy nuclei of the same total energy.  The measured depth of 
shower maximum increases more rapidly than expected from model
calculations, indicating an increasing fraction of protons as energy
increases.  A similar change was observed in analysis of
the Fly's Eye data~\cite{FlyEye} but at a higher energy (above $10^{18}$~eV).
It was recognized at the time that such a change of composition could
be a signature of the transition from a population of 
galactic cosmic rays to an extra-galactic population.
In view of the more recent data, the signature of a transition
appears to occur at a lower energy (below $10^{18}$~eV).
A discussion and references to cosmological scenarios in
which the transition to extra-galactic cosmic rays occurs
at relatively low energy is given by Hillas~\cite{Hillas}.

Several new experiments are planned or in operation that can
explore the energy region from the knee to the EeV region
to overlap with the threshold region of the giant arrays 
Auger~\cite{Auger} and (formerly) AGASA~\cite{AGASA}. 
The most advanced of these if KASCADE-Grande~\cite{K-G}, a large sea-level
array that includes the original KASCADE as a subarray.
IceCube is under construction as described in \S\ref{sec4}.
Telescope Array~\cite{TA} and its low-energy extension, TALE~\cite{TALE}, 
are under development in Utah.  There is a propsed array of atmospheric 
Cherenkov detectors, TUNKA~\cite{TUNKA}, in Russia, and use of the radio technique
for a large acceptance air shower detector is being explored~\cite{LOPES}.  
Understanding the transition from a galactic to an extra-galactic population
of cosmic rays is an interesting and important goal for the near future.

\noindent{\bf Acknowledgments} I thank Peter Niessen for the generating the data for
Fig.~\ref{fig3} and Michael Hillas for enlightening discussions.  
This research is supported in part by the U.S. Department of Energy
under DE-FG0291 ER 40626.


\end{document}